\documentclass{iopart}

\usepackage{graphicx}

\begin{document}

\title{
Enhanced Pair-correlation functions in the two-dimensional Hubbard model
}

\author{Takashi Yanagisawa} 

\address{Electronics and Photonics Research Institute, 
National Institute of Advanced Industrial Science and Technology (AIST),
Central 2, 1-1-1 Umezono, Tsukuba, Ibaraki 305-8568, Japan
}

\ead{t-yanagisawa@aist.go.jp}

\begin{abstract}
In this study we have computed the pair correlation functions in the
two-dimensional Hubbard model using a quantum Monte Carlo method.
We employ a new diagonalization algorithm in quantum Monte Carlo method
which is free from the negative sign problem.
We show that
the d-wave pairing correlation function is indeed enhanced slightly for 
the positive
on-site Coulomb interaction $U$ when doping away from the half-filling.
When the system size becomes large, the pair correlation function $P_d$ is
increased for $U>0$ compared to the non-interacting case, while $P_d$ is 
suppressed for $U>0$
when the system size is small.
The enhancement ratio $P_d[U]/P_d[U=0]$ will give a criterion on the
existence of superconductivity.
The ratio $P_d[U]/P_d[U=0]$ increases almost linearly
$\propto L$ as the system size
$L\times L$ is increased.
This increase is a good indication of an existence of superconducting phase in
the two-dimensional Hubbard model.
There is, however, no enhancement of pair correlation 
functions at half-filling, which indicates the absence of superconductivity
without hole doping.

\end{abstract}

\pacs{74.20.-z, 71.10.Fd, 75.40.Mg}

\maketitle

\section{Introduction}
Strongly correlated electron systems have been studied intensively
in relation to high-temperature superconductivity (SC).
High-temperature superconductors\cite{dag94,sca90,and97,mor00} are
known as a typical correlated electron system.
Recently, the mechanism of 
superconductivity in high-temperature superconductors 
has been extensively studied using various two-dimensional 
(2D) models of electronic interactions.  
Among them the 2D Hubbard model\cite{hub63} is the simplest and most 
fundamental model.
This model has been studied intensively using numerical tools, such as the 
quantum Monte Carlo (QMC) method
\cite{hir83,hir85,sor88,whi89,ima89,sor89,loh90,mor91,fur92,mor92,fah91,zha97,
zha97b,kas01,yan98,yan07},
and the variational Monte Carlo (VMC) 
method\cite{yok87,gro87,nak97,yam98,koi99,yan01,yan02,yan03,yan05,miy04,luc12,yok12}.

The Quantum Monte Carlo method is a numerical method employed to 
simulate the behavior of correlated electron systems. 
It is well known, however, that there are significant issues associated with
the application of the QMC method.
The most important one is that the standard Metropolis
(or heat bath) algorithm is associated with the negative sign problem.  
In past studies workers have investigated the possibility of eliminating
the negative sign 
problem\cite{fah91,zha97,kas01,yan07}.
 
In this paper we adopt an optimization scheme which is based on 
diagonalization  Quantum Monte Carlo (QMD) method\cite{yan07} (a bosonic
version was developed in Ref.\cite{miz86}),
as well as the Metropolis Quantum Monte Carlo method
(called the Metropolis QMC in this paper).
In general, and as in this study, the ground-state wave function is defined as
\begin{equation}
\psi= e^{-\tau H}\psi_0,
\end{equation}
where $H$ is the Hamiltonian and $\psi_0$ is the initial one-particle
state such as the Fermi sea.
In the QMD method this wave function is written as a 
linear combination of the basis states, generated using the auxiliary 
field method based on the Hubbard-Stratonovich transformation; that is  
\begin{equation}
\psi= \sum_mc_m\phi_m,
\end{equation}
where $\phi_m$ are basis functions.
In this work we have assumed a subspace with $N_{states}$ basis wave functions.
From the variational principle, the coefficients $\{c_m\}$ are
determined from the diagonalization of the Hamiltonian, 
to obtain the lowest energy state in the selected subspace $\{\phi_m\}$.  
Once the $c_m$ coefficients are determined, the ground-state energy and 
other quantities
are calculated using this wave function.
If the expectation values are not highly sensitive to the number of basis
states, we can obtain the correct expectation values using an extrapolation
in terms of the basis states in the limit $N_{states}\rightarrow\infty$.

Whether the 2D Hubbard model can account for high-temperature 
superconductivity is an
important question in the study of high-temperature superconductors.
In correlated electron systems, there is an interesting phenomenological
correlation between the maximum $T_c$ and the transfer integral $t$:
\begin{equation}
k_BT_c \simeq 0.1t/(m^*/m).
\end{equation}
$m^*/m$ indicates the mass enhancement factor and $t_{eff}\equiv t/(m^*/m)$ 
is the effective transfer integral.
By adopting $t\sim 0.5$eV\cite{fei96} and $m^*/m\sim 5$,
this formula applies to high-$T_c$ cuprates with $T_c\sim 100$K.
As the electron becomes heavier, $T_c$ is lowered (in accordance with
the lowering of $T_c$ in the underdoped region).
We can choose $t\sim 0.1$eV and $m*/m\sim 2$ for iron pnictides to
give $T_c\sim 50$K.
This formula strongly suggests that high-temperature superconductivity
originates from the
electron correlation, not from the electron-phonon interaction.

Most of QMC method results do not support superconductivity, although the results
of VMC method with the Gutzwiller ansatz indicates the stable d-wave pairing
state for large $U$.
The computations of the pair-field susceptibility suggest the existence of
the Kosterlitz-Thouless transition in the 2D Hubbard model indicating
superconducting transition in real 3D systems\cite{mai05,yan10}.
The perturbative and Random phase approximation (RPA) calculations also 
support superconductivity with
anisotropic pairing symmetry\cite{sca86,bic89,hlu99,kon01,yan08}.
In contrast, the pair correlation functions obtained by a QMC method\cite{zha97b}
are extremely suppressed
for the intermediate values of $U$.
This result suggests that superconductivity is impossible in the 2D
Hubbard model.
The objective of this paper is to compute pair correlation functions 
and clarify this discrepancy using
a new QMC method with employing the diagonalization scheme\cite{yan07}.
We show that the pair correlation function is indeed enhanced at doping.

\section{Model and the Wave function}

\subsection{Hamiltonian}

The Hamiltonian is the Hubbard model containing on-site Coulomb 
repulsion and is written as

\begin{eqnarray}
H&=&-\sum_{ij\sigma}t_{ij}(c^{\dag}_{i\sigma}c_{j\sigma}+h.c.)
+ U\sum_jn_{j\uparrow}n_{j\downarrow},
\label{hamil}
\end{eqnarray}
where $c^{\dag}_{j\sigma}$ ($c_{j\sigma}$) is the creation 
(annihilation) operator of an electron with spin $\sigma$
at the $j$-th site and $n_{j\sigma}=c^{\dag}_{j\sigma}c_{j\sigma}$.  
$t_{ij}$ is the transfer energy between the sites $i$ and $j$.
$t_{ij}=t$ for the nearest-neighbor bonds and $t_{ij}=-t'$ for the next
nearest-neighbor bonds. For all other cases $t_{ij}=0$.
$U$ is the on-site Coulomb energy. 
The number of sites is $N$ and
the linear dimension of the system is denoted as $L$, i.e. $N=L^2$.
The energy unit is given by $t$ and
the number of electrons is denoted as $N_e$.

\subsection{Quantum Monte Carlo method - Metropolis algorithm}

In a Quantum Monte Carlo simulation, the ground state wave function is
\begin{equation}
\psi= {\rm e}^{-\tau H}\psi_0 ,
\label{psi}
\end{equation}
where $\psi_0$ is the initial one-particle state represented by a Slater 
determinant.  
For large $\tau$, ${\rm e}^{-\tau H}$ will project out the ground state from
$\psi_0$.
We write the Hamiltonian as $H=K+V$ where
K and V are the kinetic and interaction terms of the Hamiltonian in 
Eq.(\ref{hamil}), respectively. 
The wave function in Eq.(\ref{psi}) is written as
\begin{equation}
\psi= ({\rm e}^{-\Delta\tau (K+V)})^m \psi_0 \approx 
({\rm e}^{-\Delta\tau K}{\rm e}^{-\Delta\tau V})^m \psi_0 ,
\end{equation}
for $\tau=\Delta\tau\cdot m$.
Using the Hubbard-Stratonovich transformation\cite{hir83,bla81}, we have
\begin{equation}
{\rm exp}(-\Delta\tau Un_{i\uparrow}n_{i\downarrow})
= \frac{1}{2}\sum_{s_i=\pm 1}{\rm exp}( 2as_i(n_{i\uparrow}
-n_{i\downarrow})
-\frac{1}{2}U\Delta\tau(n_{i\uparrow}+n_{i\downarrow}) ),
\end{equation}
for $({\rm tanh}a)^2={\rm tanh}(\Delta\tau U/4)$ or
${\rm cosh}(2a)={\rm e}^{\Delta\tau U/2}$.
The wave function is expressed as a summation of the one-particle Slater
determinants
over all the configurations of the auxiliary fields 
$s_j=\pm 1$.
The exponential operator is expressed as\cite{bla81}
\begin{eqnarray}
({\rm e}^{-\Delta\tau K}{\rm e}^{-\Delta\tau V})^m
&=& \frac{1}{2^{Nm}}\sum_{\{s_i(\ell)\}}\prod_{\sigma}B_m^{\sigma}(s_i(m))
\nonumber\\
&\times& B_{m-1}^{\sigma}(s_i(m-1))\cdots B_1^{\sigma}(s_i(1)),\nonumber\\
\end{eqnarray}
where we have defined
\begin{equation}
B_{\ell}^{\sigma}(\{s_i(\ell)\})={\rm e}^{-\Delta\tau K_{\sigma}}
{\rm e}^{-V_{\sigma}(\{s_i(\ell)\})},
\label{bmat}
\end{equation}
for
\begin{equation}
V_{\sigma}(\{s_i\})= 2a\sigma\sum_i s_in_{i\sigma}-\frac{1}{2}
U\Delta\tau \sum_in_{i\sigma},
\end{equation}
\begin{equation}
K_{\sigma}=-\sum_{ij}t_{ij}(c_{i\sigma}^{\dag}c_{j\sigma}+h.c.).
\end{equation}
The ground-state wave function is
\begin{equation}
\psi= \sum_nc_n\phi_n ,
\label{wf}
\end{equation}
where $\phi_n$ is a Slater determinant corresponding to a configuration 
$\{s_i(\ell)\}$ ($i=1,\cdots,N; \ell=1,\cdots,m$)
of the auxiliary fields:
\begin{eqnarray}
\phi_n&=& \prod_{\sigma}B_m^{\sigma}(s_i(m))\cdots B_1^{\sigma}(s_i(1))\psi_0
\nonumber\\
&\equiv& \phi_n^{\uparrow}\phi_n^{\downarrow}.
\end{eqnarray}
The coefficients $c_n$ are constant real numbers: $c_1=c_2=\cdots$.
The initial state $\psi_0$ is a one-particle state.
The matrix of $V_{\sigma}(\{s_i\})$ is a
diagonal matrix given as
\begin{equation}
V_{\sigma}(\{s_i\})={\rm diag}(2a\sigma s_1-U\Delta\tau/2,\cdots,
2a\sigma s_N-U\Delta\tau/2).
\end{equation}
The matrix elements of $K_{\sigma}$ are
\begin{eqnarray}
(K_{\sigma})_{ij}&=& -t~~~i,j~ {\rm are~ nearest~ neighbors}\nonumber\\
&=& 0~~~{\rm otherwise}.
\end{eqnarray}
$\phi_n^{\sigma}$ is an $N\times N_{\sigma}$ matrix given by the product
of the matrices ${\rm e}^{-\Delta\tau K_{\sigma}}$, ${\rm e}^{V_{\sigma}}$
and $\psi_0^{\sigma}$.
The inner product is thereby calculated as a determinant\cite{zha97},
\begin{equation}
\langle\phi_{\ell}^{\sigma}\phi_n^{\sigma}\rangle=
{\rm det}(\phi_{\ell}^{\sigma\dag}\phi_n^{\sigma}).
\end{equation}
The expectation value of the quantity $Q$ is evaluated as
\begin{equation}
\langle Q\rangle = \frac{\sum_{\ell n}\langle\phi_{\ell} Q\phi_n\rangle}
{\sum_{\ell n}\langle\phi_{\ell}\phi_n\rangle}.
\label{qexpe}
\end{equation}

$P_{\ell n}\equiv {\rm det}(\phi_{\ell}^{\sigma}\phi_n^{\sigma})
{\rm det}(\phi_{\ell}^{-\sigma}\phi_n^{-\sigma})$
can be regarded as the weighting factor to obtain the Monte Carlo samples.
Since this quantity is not necessarily positive definite, the weighting factor
should be $|P_{\ell n}|$; the resulting relationship is,
\begin{eqnarray}
\langle Q_{\sigma}\rangle&=& \sum_{\ell n}P_{\ell n}\langle Q_{\sigma}\rangle_{\ell n}
/\sum_{\ell n}P_{\ell n}\nonumber\\
&=& \sum_{\ell n}|P_{\ell n}|sign(P_{\ell n})\langle Q_{\sigma}\rangle_{\ell n}
/\sum_{\ell n}|P_{\ell n}|sign(P_{\ell n})\nonumber\\
\end{eqnarray}
where
$sign(a)=a/|a|$ and
\begin{equation}
\langle Q_{\sigma}\rangle_{\ell n}=\frac{\langle\phi_{\ell}^{\sigma}Q_{\sigma}
\phi_n^{\sigma}\rangle}{\langle\phi_{\ell}^{\sigma}\phi_n^{\sigma}\rangle}.
\end{equation} 
This relation can be evaluated using a Monte Carlo procedure if an 
appropriate algorithm, such as the Metropolis or heat bath method, is 
employed\cite{bla81}.
The summation can be evaluated using appropriately defined Monte Carlo
samples,
\begin{equation}
\langle Q_{\sigma}\rangle= \frac{ \frac{1}{n_{MC}}\sum_{\ell n}sign(P_{\ell n})
\langle Q_{\sigma}\rangle_{\ell n}}{\frac{1}{n_{MC}}\sum_{mn}sign(P_{\ell n})},
\label{qqmc}
\end{equation}
where $n_{MC}$ is the number of samples.
The sign problem is an issue if the summation of $sign(P_{\ell n})$
vanishes within statistical errors.  In this case it is indeed impossible
to obtain definite expectation values.

\subsection{Quantum Monte Carlo method - Diagonalization algorithm}

Quantum Monte Carlo diagonalization (QMD) is a method for the evaluation of
$\langle Q_{\sigma}\rangle$ without {\em the negative sign problem}.
The configuration space of the probability $\|P_{mn}\|$ in Eq.(\ref{qqmc}) 
is generally very strongly peaked.
The sign problem lies in the distribution of $P_{mn}$ in the configuration
space.
It is important to note that the distribution of the basis functions $\phi_m$
($m=1,2,\cdots$) is uniform since $c_m$ are constant numbers: $c_1=c_2=\cdots$.
In the subspace $\{\phi_m\}$, selected from all configurations of auxiliary
fields, the right-hand side of Eq.(\ref{qexpe}) can be determined.
However, the large number of basis states required to obtain accurate
expectation values is beyond the current storage capacity of computers.
Thus we use the variational principle to obtain the expectation values.

From the variational principle,
\begin{equation}
\langle Q\rangle = \frac{\sum_{mn}c_mc_n\langle\phi_m Q\phi_n\rangle}
{\sum_{mn}c_mc_n\langle\phi_m\phi_n\rangle},
\end{equation}
where $c_m$ ($m=1,2,\cdots$) are variational parameters.
In order to minimize the energy
\begin{equation}
E = \frac{\sum_{mn}c_mc_n\langle\phi_m H\phi_n\rangle}
{\sum_{mn}c_mc_n\langle\phi_m\phi_n\rangle},
\end{equation}
the equation $\partial E/\partial c_n=0$ ($n=1,2,\cdots$) is solved for,
\begin{equation}
\sum_m c_m\langle\phi_n H\phi_m\rangle-E\sum_m c_m\langle\phi_n\phi_m\rangle
=0.
\end{equation}
If we set
\begin{equation}
H_{mn}=\langle\phi_m H\phi_n\rangle,
\end{equation}
\begin{equation}
A_{mn}=\langle\phi_m\phi_n\rangle,
\end{equation}
the eigen equation is
\begin{equation}
Hu=EAu,
\end{equation}
for $u=(c_1,c_2,\cdots)^t$.
Since $\phi_m$ ($m=1,2,\cdots$) are not necessarily orthogonal,
$A$ is not a diagonal matrix.
We diagonalize the Hamiltonian $A^{-1}H$, and then
calculate the expectation values of correlation functions 
with the ground state eigenvector;  
in general $A^{-1}H$ is not a symmetric matrix.

In order to optimize the wave function we must increase the number of
basis states $\{\phi_m\}$.
This can be simply accomplished through random sampling.
For systems of small sizes and small $U$, we can evaluate the expectation
values from an extrapolation of the basis of randomly generated states.
The number of basis states is about 2000 when the system size is small.
For systems $8\times 8$ and $10\times 10$, the number of states in 
increased up to about 10000.

  In Quantum Monte Carlo simulations an extrapolation is performed to obtain 
the expectation values for the ground-state wave function.  
The variance method has been proposed in
variational and Quantum Monte Carlo simulations, where the extrapolation is
performed as a function of the variance.
An advantage of the variance method lies is that
linearity is expected in some
cases\cite{sor01,kas01}:
\begin{equation}
\langle Q\rangle-Q_{exact}\propto v,
\end{equation}
where $v$ denotes the variance defined as
\begin{equation}
v= \frac{\langle (H-\langle H\rangle)^2\rangle}{\langle H\rangle^2}
\label{vari}
\end{equation}
and $Q_{exact}$ is the expected exact value of the quantity $Q$.

\begin{figure}
\begin{center}
\begin{tabular}{c}

\begin{minipage}{0.5\hsize}
\begin{center}
\includegraphics[width=6cm]{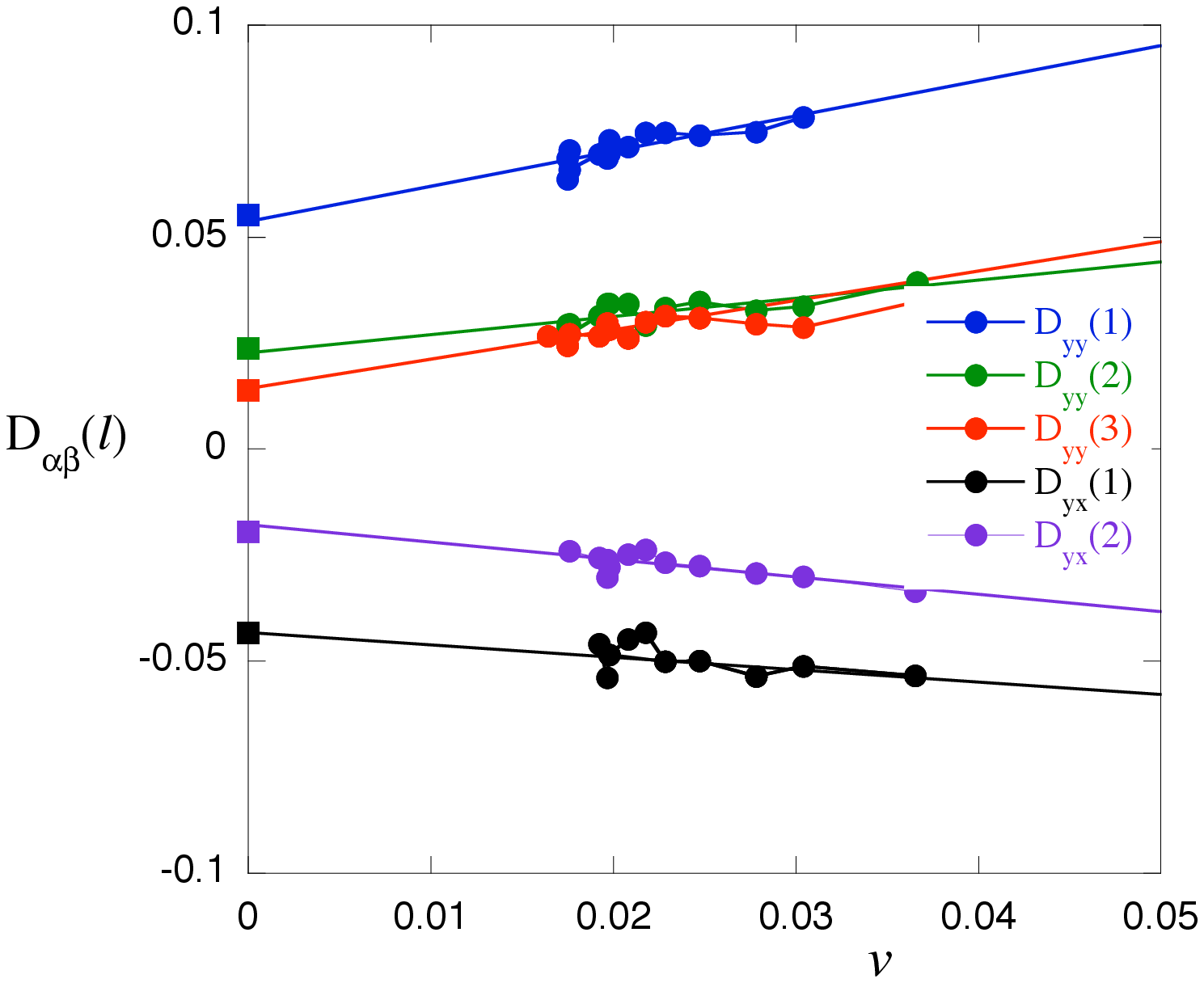}
\hspace{1cm} (a)
\end{center}
\label{4x3U4qmd}
\end{minipage}

\begin{minipage}{0.5\hsize}
\begin{center}
\includegraphics[width=6cm]{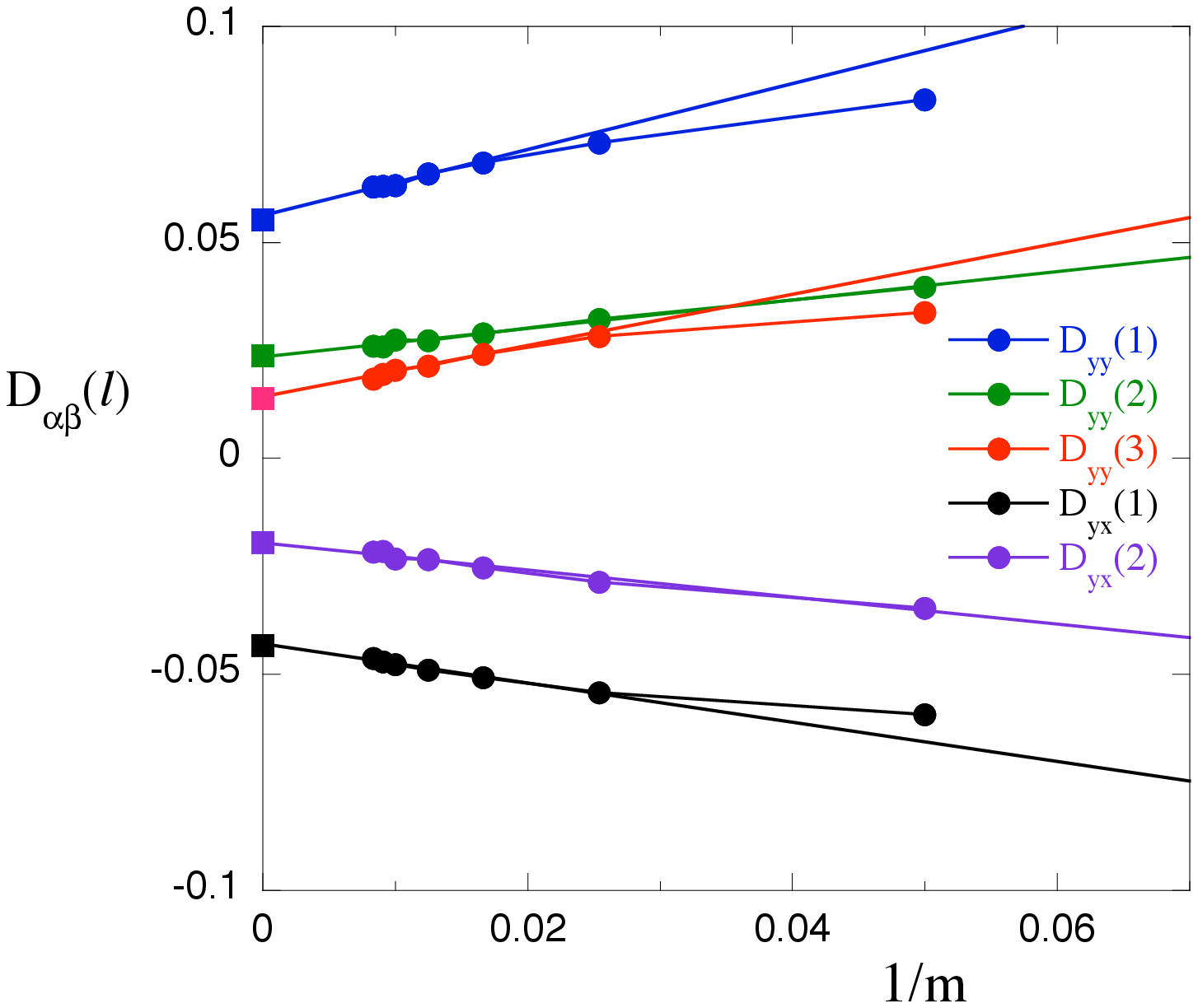}
\hspace{1cm} (b)
\end{center}
\label{4x3U4qmc}
\end{minipage}

\end{tabular}
\caption{
Pair correlation function $D_{yy}(\ell)$ and $D_{yx}(\ell)$ 
for $4\times 3$, $U=4$ and $N_e=10$
obtained by the diagonalization quantum Monte Carlo method (a) and the
Metropolis quantum Monte Carlo method (b).
The square are the exact results obtained by the exact diagonalization method.
In (a) the data fit using a straight line using the least-square method
as the variance is reduced.
We started with $N_{states}=100$ (first solid circles) and then increase 
up to 2000.
}
\end{center}
\end{figure}

%\begin{figure}
%\includegraphics[width=10cm]{4x3U4qmd}
%\caption{
%Pair correlation function $D_{yy}(\ell)$ for $4\times 3$, $U=4$ and $N_e=10$
%obtained by the quantum Monte Carlo diagonalization.
%The square are the exact results.
%The data fit using a straight line using the least-square method
%as the variance is reduced.
%We started with $N_{states}=100$ (first solid circle) and then increase 
%up to 2000.
%}
%\label{4x3U4qmd}
%\end{figure}

%\begin{figure}
%\includegraphics[width=10cm]{4x3U4qmc}
%\caption{
%Pair correlation function $D_{yy}(\ell)$ for $4\times 3$, $U=4$ and $N_e=10$
%obtained by the Metropolis quantum Monte Carlo method.
%The square are the exact results.
%}
%\label{4x3U4qmc}
%\end{figure}

%\begin{figure}
%\includegraphics[width=10cm]{30x2U2qmd}
%\caption{
%Pair correlation function $D_{yy}(\ell)$ as a function of the energy
%variance $v$ for $30\times 2$, $U=2$ and $N_e=48$
%obtained by the diagonalization quantum Monte Carlo method.
%We set the open boundary condition.
%From the top, $\ell=(1,0)$, (2,0),(5,0), (4,0), (3,0) and (6,0).
%}
%\label{30x2U2qmd}
%\end{figure}

\begin{figure}
\begin{center}
\begin{tabular}{c}

\begin{minipage}{0.5\hsize}
\begin{center}
\includegraphics[width=6cm]{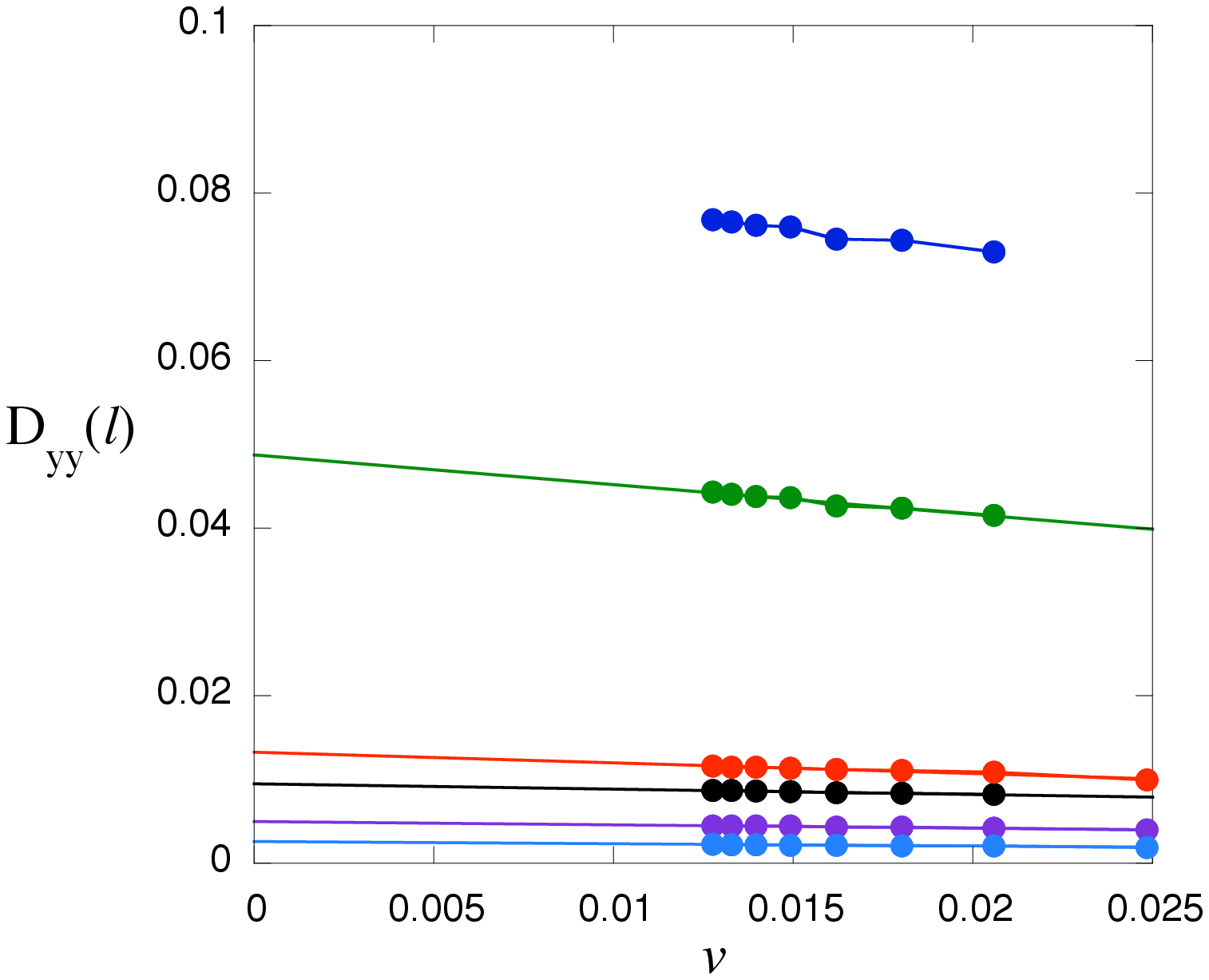}
\hspace{1cm} (a)
\end{center}
\label{30x2U4qmd}
\end{minipage}

\begin{minipage}{0.5\hsize}
\begin{center}
\includegraphics[width=6cm]{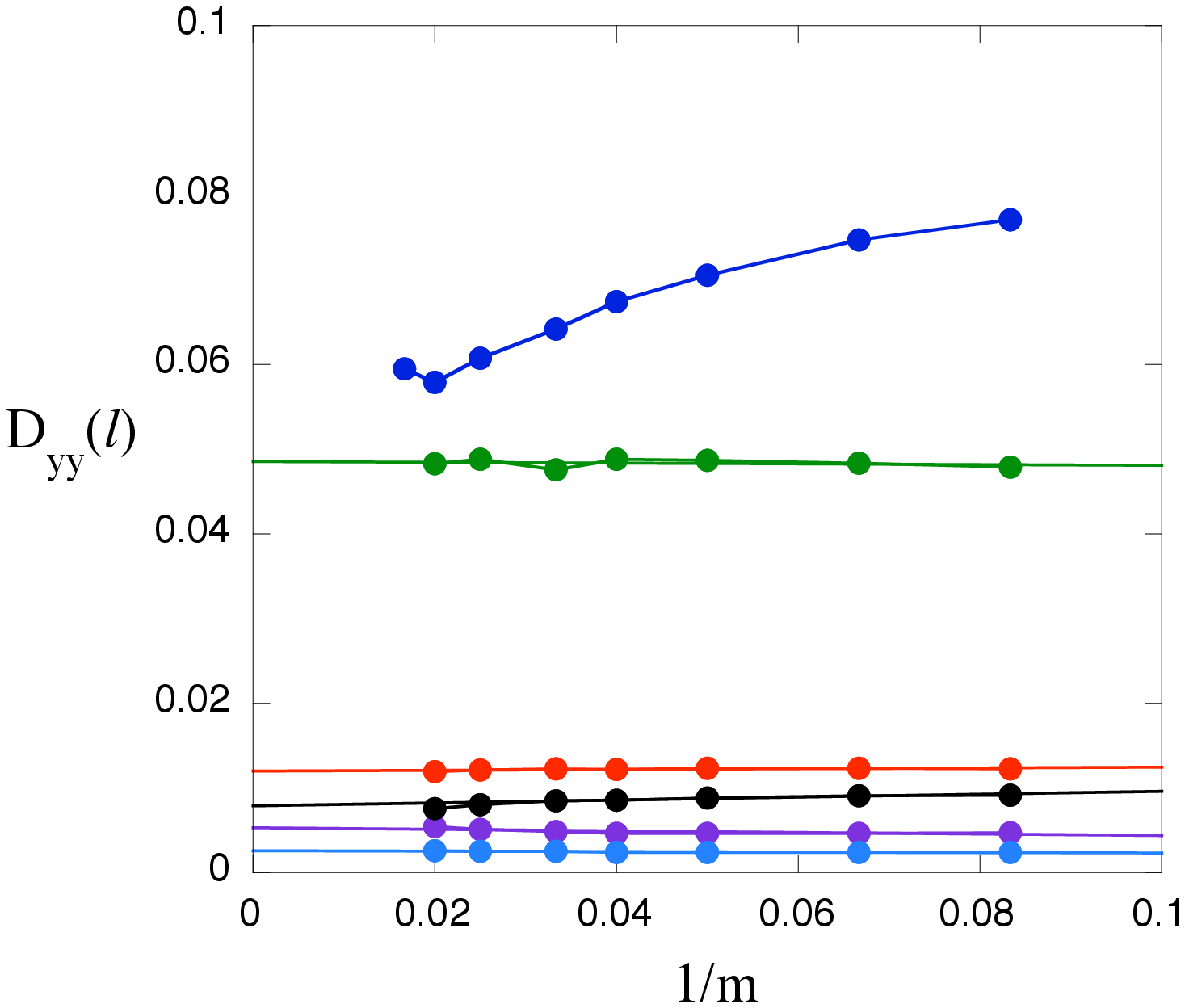}
\hspace{1cm} (b)
\end{center}
\label{30x2U4qmc}
\end{minipage}

\end{tabular}
\caption{
Pair correlation function $D_{yy}(\ell)$ as a function of the energy
variance $v$ in (a) and $1/m$ in (b) for $30\times 2$, $U=4$ and $N_e=48$.
We used (a) the diagonalization quantum Monte Carlo method and (b) the
Metropolis quantum Monte Carlo method.
We set the open boundary condition.
From the top, $\ell=(1,0)$, (2,0),(5,0), (4,0), (3,0) and (6,0).
}
\end{center}
\end{figure}

%\begin{figure}
%\includegraphics[width=10cm]{30x2U4qmd}
%\caption{
%Pair correlation function $D_{yy}(\ell)$ as a function of the energy
%variance $v$ for $30\times 2$, $U=4$ and $N_e=48$
%obtained by the diagonalization quantum Monte Carlo method.
%We set the open boundary condition.
%From the top, $\ell=(1,0)$, (2,0),(5,0), (4,0), (3,0) and (6,0).
%}
%\label{30x2U4qmd}
%\end{figure}

%\begin{figure}
%\includegraphics[width=10cm]{30x2U4qmc}
%\caption{
%Pair correlation function $D_{yy}(\ell)$ as a function of $1/m$
%for $30\times 2$, $U=4$ and $N_e=48$
%obtained by the Metropolis quantum Monte Carlo method.
%We set the open boundary condition.
%From the top, $\ell=(1,0)$, (2,0),(5,0), (4,0), (3,0) and (6,0).
%}
%\label{30x2U4qmc}
%\end{figure}

\section{Pair correlation functions}

In this section, we present the results obtained by the QMC and QMD methods.

\subsection{Comparison of two methods}

The pair correlation function $D_{\alpha\beta}$ is defined by
\begin{equation}
D_{\alpha\beta}(\ell)= \langle \Delta_{\alpha}^{\dag}(i+\ell)
\Delta_{\beta}(i)\rangle,
\end{equation}
where $\Delta_{\alpha}(i)$, $\alpha=x,y$, denote the annihilation operators
of the singlet electron pairs for the nearest-neighbor sites:
\begin{equation}
\Delta_{\alpha}(i)= c_{i\downarrow}c_{i+\hat{\alpha}\uparrow}
-c_{i\uparrow}c_{i+\hat{\alpha}\downarrow}.
\end{equation}
Here $\hat{\alpha}$ is a unit vector in the $\alpha(=x,y)$-direction.
We consider the correlation function of d-wave pairing:
\begin{equation}
P_d(\ell)= \langle \Delta_d(i+\ell)^{\dag}\Delta_d(i)\rangle,
\end{equation}
where
\begin{equation}
\Delta_d(i)= \Delta_x(i)+\Delta_{-x}(i)-\Delta_y(i)-\Delta_{-y}(i).
\end{equation}
$i$ and $i+\ell$ denote sites on the lattice.

We show how the pair correlation function is evaluated in quantum
Monte Carlo methods.
We show the pair correlation functions $D_{yy}$ and $D_{yx}$ on the lattice 
$4\times 3$ in Fig.1.
The boundary condition is open in the 4-site direction and is periodic
in the other direction.
An extrapolation is performed as a function of $1/m$ in the QMC
method with Metropolis algorithm and as a function of the energy 
variance $v$ in the QMD method with diagonalization. 
We keep $\Delta\tau$ a small constant $\simeq 0.02\sim 0.05$ and
and increase $\tau=\Delta\tau\cdot m$, where $m$ is the division number
$m$ of the wave function $\psi$ in eq.(5).
In the Metropolis QMC method, we calculated averages over $5\times 10^5$
Monte Carlo steps.
The exact values were obtained by using the exact diagonalization method.
Two methods give consistent results as shown in figures.
All the $D_{yy}(\ell)$ and
$D_{yx}(\ell)$ are suppressed on $4\times 3$ as $U$ is increased.
In general, the pair correlation functions are suppressed in small
systems.

In Fig.2, we show the inter-chain pair correlation function
$D_{yy}(\ell)$ as a function of $1/m$ (b) and the energy
variance (a) for the ladder model $30\times 2$.
We use the open boundary condition.
The boundary condition is not important for our purpose to check the
consistency between QMC and QMD mthods.
The number of electrons is $N_e=48$, and the strength of the Coulomb interaction
is $U=4$.
$\Delta_y(i)$ indicates the electron pair along the rung, and $D_{yy}(\ell)$
is the expectation value of the parallel movement of the pair along the
ladder.
The results obtained by two methods are in good agreement except
$\ell=(1,0)$ (nearest-neighbor correlation).

\begin{figure}
\begin{center}
\includegraphics[width=8cm]{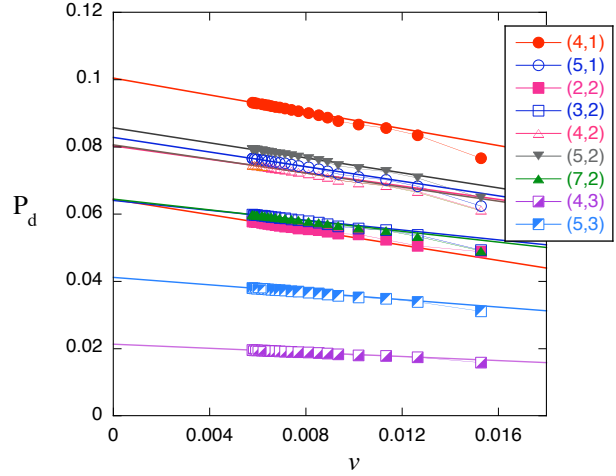}
\caption{
Pair correlation function $P_{d}$ as a function of the energy variance $v$
on $8\times 8$ lattice.  $U=3$, $t'=-0.2$ and the electron number is 
$N_e=54$.
We have shown $P_d(\ell)=\langle\Delta_d(i+\ell)^{\dag}\Delta(i)\rangle$ 
for $\ell=(m,n)-i$ and $i=(1,1)$, where $(m,n)$ are
shown in the figure.
}
\label{64-54U3}
\end{center}
\end{figure}

\begin{figure}
\begin{center}
\begin{tabular}{c}

\begin{minipage}{0.5\hsize}
\begin{center}
\includegraphics[width=6cm]{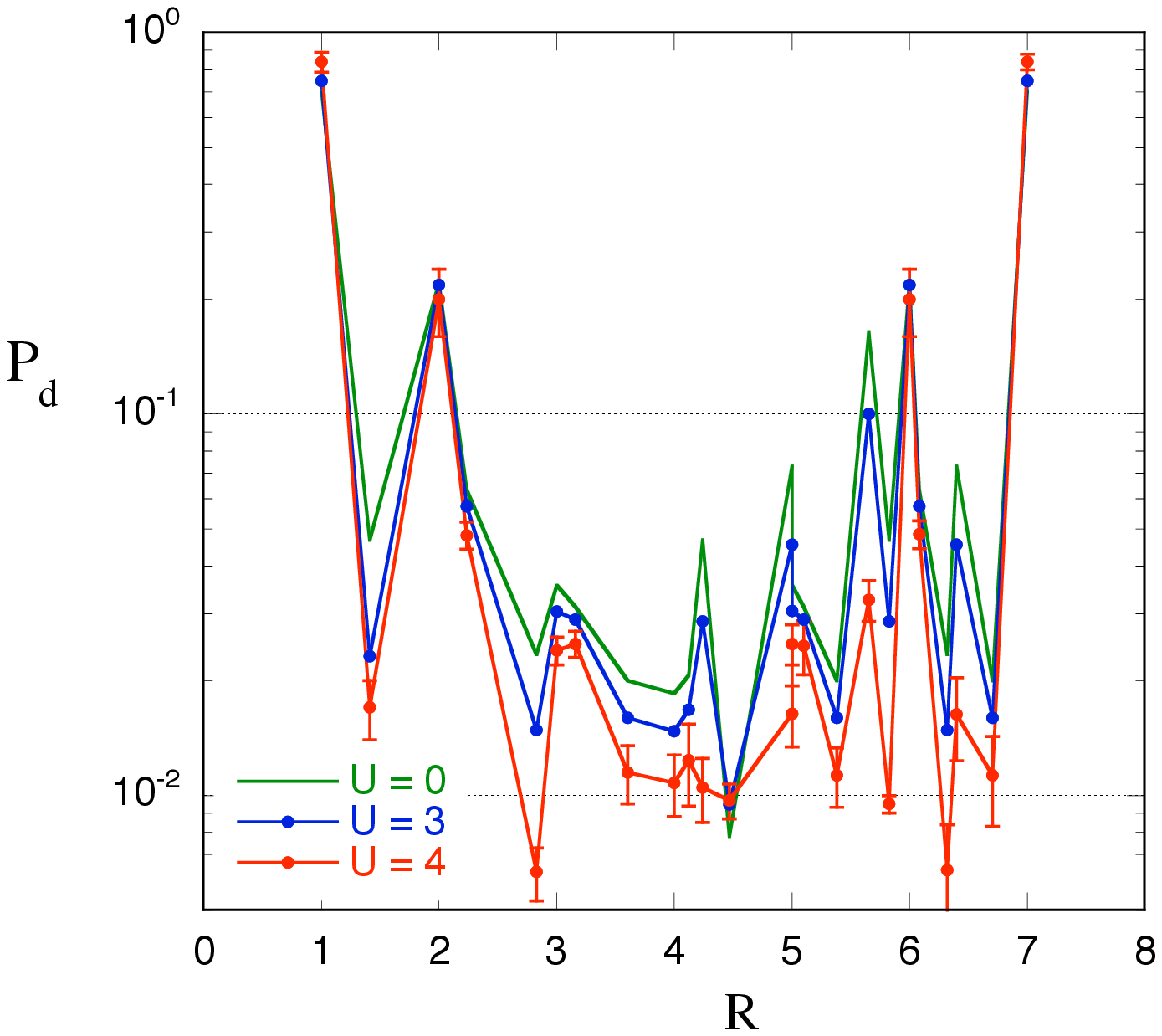}
\hspace{1cm} (a)
\end{center}
\label{64-64sc-r}
\end{minipage}

\begin{minipage}{0.5\hsize}
\begin{center}
\includegraphics[width=6cm]{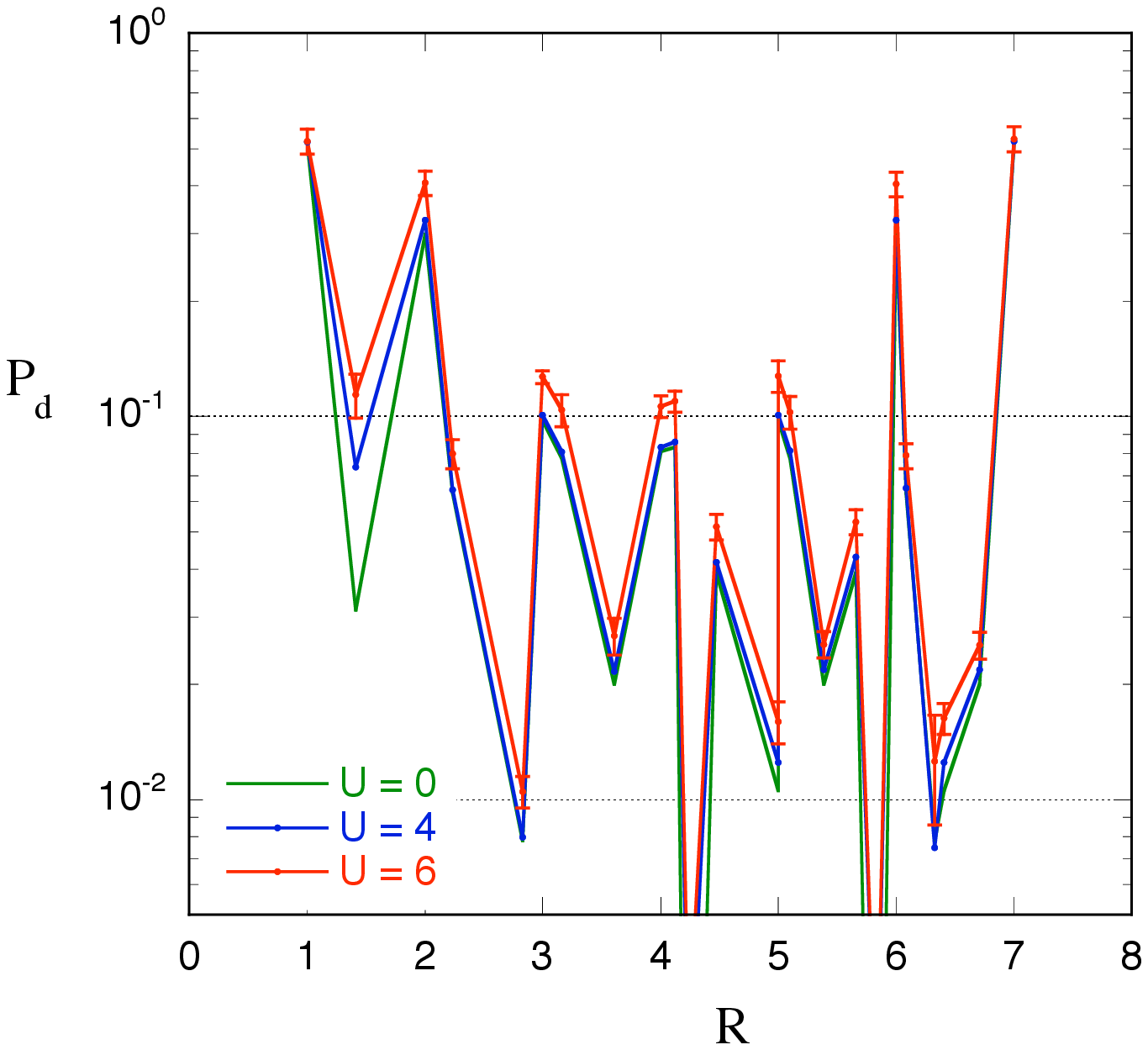}
\hspace{1cm} (b)
\end{center}
\label{64-54sc-r}
\end{minipage}

\end{tabular}
\caption{
Pair correlation function $P_{d}$ as a function of the distance $R=|\ell|$
on $8\times 8$ lattice for (a) the half-filled case $N_e=64$ and (b) $N_e=54$.  
We set $t'=0.0$ and $U=0$, 3 and 4 for (a) and $t'=-0.2$ and $U=0$, 4 and
6 for (b).
To lift the degeneracy of electron configurations at the Fermi energy
in the half-filled case,
we included a small staggered magnetization $\sim 10^{-4}$ in the
initial wave function $\psi_0$. 
}
\end{center}
\end{figure}

%\begin{figure}
%\includegraphics[width=10cm]{64-56sc-r}
%\caption{
%Pair correlation function $P_{d}$ as a function of the distance $R=|\ell|$
%on $8\times 8$ lattice for $N_e=56$.  
%We set $t'=-0.2$, and $U=0$, 4 and 6. This is the open shell case.
%}
%\label{64-56sc-r}
%\end{figure}

\begin{figure}
\begin{center}
\includegraphics[width=8cm]{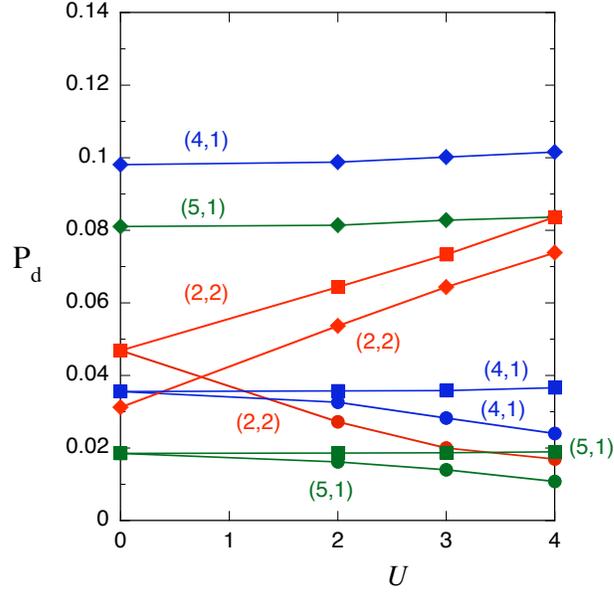}
\caption{
Pair correlation function $P_{d}$ as a function of $U$
on $8\times 8$ lattice.  $t'=-0.2$ for $N_e=54$ (diamonds), 
and $t'=0$ for $N_e=50$ (squares) 
and $N_e=64$ (circles).
We have shown $P_d(\ell)=\langle\Delta_d(i+\ell)^{\dag}\Delta(i)\rangle$
for $\ell=(m,n)-i$ and $i=(1,1)$, where $(m,n)$ are
shown in the figure.
}
\label{64-SC-U}
\end{center}
\end{figure}

\begin{figure}
\begin{center}
\includegraphics[width=8cm]{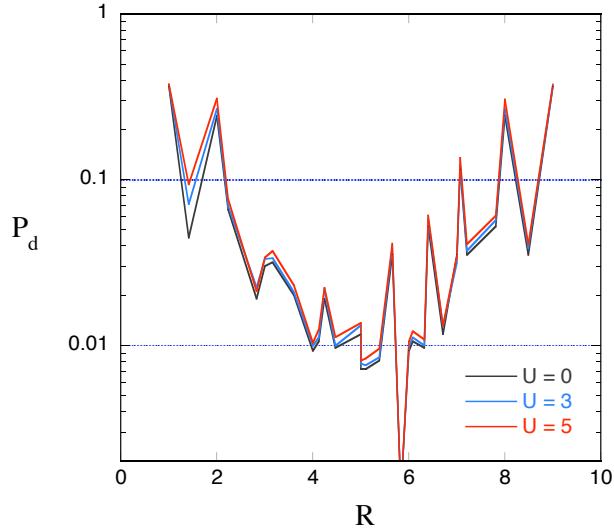}
\caption{
Pair correlation function $P_{d}$ as a function of the distance $R=|\ell|$
on $10\times 10$ lattice for $N_e=82$ and $t'=-0.2$.  
The strength of the Coulomb interaction is $U=0$, 3 and 5.
}
\label{100-82sc-r}
\end{center}
\end{figure}

\begin{figure}
\begin{center}
\includegraphics[width=8cm]{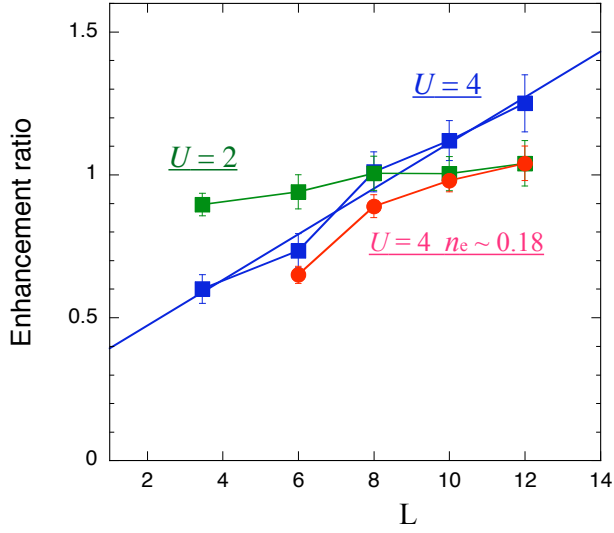}
\caption{
Enhancement ratio of pair correlation function $P_d|_U/P_d|_{U=0}$ 
as a function of the linear system size $L$ for $U=4$ and $U=2$.
The electron density $n_e$ is about 0.8: $n_e\sim 0.8$ for squares.
The data for $U=4$ and $n_e\sim 0.18$ are also shown by circles.
}
\label{enhance}
\end{center}
\end{figure}

\begin{figure}
\begin{center}
\includegraphics[width=8cm]{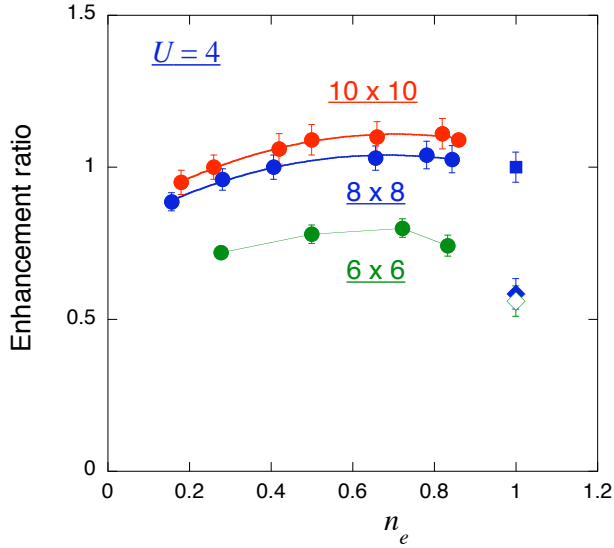}
\caption{
Enhancement ratio of pair correlation function $P_d|_U/P_d|_{U=0}$ 
as a function of the electron density $n_e$.
We adopt $t'=-0.2$ and $U=4$.  For the half-filled case, the diamonds show
that for $t'=0$ on $8\times 8$ lattice (solid diamond) and $6\times 6$
lattice (open diamond).
The square is for $t'=-0.2$ on $8\times 8$ and $10\times 10$ where
there is no enhancement.
}
\label{enhance2}
\end{center}
\end{figure}

\subsection{Pair correlation in 2D Hubbard model}

We present the results for pair correlation in the two-dimensional
Hubbard model.
In this section we show the results using the diagonalization
QMC method because the Metropolis QMC method has a negative sign problem.
We first examine the $8\times 8$ lattice.
The $P_d$ was estimated by an extrapolation as a function of the
variance $v$, as shown in Fig.3, where the computations were carried
out on $8\times 8$ lattice with $U=3$, $t'=-0.2$ and $N_e=54$.
The extrapolation was successfully performed for $8\times 8$.

We consider the half-filled case with $t'=0$; in this case the 
antiferromagnetic correlation is dominant over the superconductive
pairing correlation and thus the pairing correlation function is
suppressed as the Coulomb repulsion $U$ is increased. 
The Fig.4(a) exhibits the d-wave pairing correlation function
$P_d$ on $8\times 8$ lattice as a function of the distance.
The $P_d$ is suppressed due to the on-site Coulomb interaction,
as expected.  Its reduction is, however, not so considerably large
compared to previous QMC studies \cite{zha97b} where the pairing
correlation is almost annihilated for $U=4$.
We then turn to the case of less than half-filling.
We show the results on $8\times 8$ with electron number
$N_e=54$.
We show $P_d$ as a function of the distance in Fig.4(b) ($N_e=54$). 
In the scale of this figure, $P_d$ for $U>0$ is almost the same
as that of the non-interacting case, and is enhanced slightly for
large $U$.
Our results indicate that the pairing correlation is not suppressed
and is indeed enhanced by the Coulomb interaction $U$, and its
enhancement is very small. 
The Fig.5 represents $P_d$ as a function of $U$ for $N_e=54$, 50 and 64.
We set $t'=0$ for $N_e=50$ and $t'=-0.2$ for $N_e=54$ so that we
have the closed shell structure in the initial function.
In the system of this size, the effect of the inclusion of $t'\neq 0$
is small.
The Fig.6 shows $P_d$ on $10\times 10$ lattice.
This also indicates that the pairing correlation function is
enhanced for $U>0$.
There is a tendency that $P_d$ is easily suppressed as the system
size becomes small.  
We estimated the enhancement ratio compared to the non-interacting
case $P_d(\ell)|_U/P_d(\ell)|_{U=0}$ at $|\ell|\sim L/2$
for $n_e\sim 0.8$ as shown in Fig.7.
This ratio increases as the system size is increased.
To compute the enhancement, we picked the sites, for example on
$8\times 8$ lattice, $\ell=(3,2)$, (4,0), (4,1), (3,3), (4,2), (4,3), (5,0),
(5,1) with $|\ell|\sim 4-5$ and evaluate the mean value.
In our computations, the ratio increases almost linearly
indicating a possibility of superconductivity.
This indicates $P_d(\ell)\sim LP_d(\ell)\sim \ell P_d(\ell)$ for $\ell\sim L$.
Because $P_d(\ell)|_{U=0}\sim 1/|\ell|^3$, we obtain
$P_d(\ell)\sim \ell P_d(\ell)\sim 1/|\ell|^2$ for $|\ell|\sim L$.
This indicatesthat the exponent of the power law is 2.
When $U=2$, the enhancement is small and is almost independent of $L$.
In the low density case, the enhancement is also suppressed being equal
to 1.
In Fig.8, the enhancement ratio is shown as a function of the electron
density $n_e$ for $U=4$.  A dome structure emerges even in small systems.
The square in Fig.8 indicates the result for the half-filled case
with $t'=-0.2$ on $8\times 8$ lattice.
This is the open shell case and causes a difficulty in computations
as a result of the degeneracy due to partially occupied electrons. 
The inclusion of $t'<0$ enhances $P_d$ compared to the case with
$t'=0$ on $8\times 8$ lattice.  $P_d$ is, however, not enhanced
over the non-interacting case at half-filling.
This also holds for $10\times 10$ lattice where the enhancement ratio
$\sim 1$.
This indicates the absence of superconductivity at half-filling.

\section{Summary}

The quest for the existence of superconducting transition in the
two-dimensional Hubbard model remains unresolved.  Pair correlation 
functions had been calculated by using QMC methods, and their results
were negative for the existence of superconductivity in many works.
The objective of this paper was to reexamine this question by
elaborating a sampling method of quantum Monte Carlo method.

We have calculated the d-wave pair correlation function $P_d$ for the 2D
Hubbard model by using the QMC method.
In the half-filled case $P_d$ is suppressed for the repulsive $U>0$, and
when doped away from half-filling $N_e<N$, $P_d$ is enhanced slightly 
for $U>0$.
It is noteworthy that the correlation function $P_d$ is
indeed enhanced and is increased as the system size increases in the 2D 
Hubbard model.
The enhancement ratio increases almost linearly $\propto L$ as the 
system size is increased, 
which an indicative of the existence of superconductivity.
Our criterion is that when the enhancement ratio as a function of the
system size $L$ is proportional to a certain power of $L$, superconductivity
will be developed.
This ratio dependes on $U$ and is reduced as $U$ is decreased.
The dependence on the band filling shows a dome structure as a function
of the electron density.
In the $10\times 10$ system, the ratio is greater than 1 in the range
$0.3 < n_e < 0.9$.  This does not immediately indicates the existence of
superconductivity.  The size dependence is important and is needed to 
obtain the doping range where superconductivity exists.
Let us also mention on superconductivity at half-filling.
Our results indicates the absence of superconductivity in the
half-filling case because there is no enhancement of pair
correlation functions.

We have compared two methods: diagonalization QMC and Metropolis QMC.
For small systems, the results obtained by two methods are quite consistent.
When the system size is large, $P_d(\ell)$ is inevitably suppressed and
almost vanishes if we use the Metropolis QMC method. 
$P_d(\ell)$ decreases as the division number $m$ increases in this method.
We wonder if this excessive suppression of $P_d(\ell)$ is true.
In fact, the correlation function $D_{yy}$ for the ladder Hubbard model
obtained by the Metropolis QMC
also shows a similar behavior when the size is increased, in contrast to
enhanced $D_{yy}$ indicated by the density-matrix renormalization 
(DMRG) method\cite{noa97}.
The results by the diagonalization QMC are consistent with those of
DMRG\cite{yan07}. 
There is a possibility that this has some relation with the negative sign.

%\section{Acknowledgments}
  We thank J. Kondo, K. Yamaji, I. Hase and S. Koikegami for helpful 
discussions.
This work was supported by Grant-in-Aid for Scientific Research from the
Ministry of Education, Culture, Sports, Science and Technology in Japan.
This work was also supported by CREST program of Japan Science and Technology
Agency (JST).
A part of numerical calculations was performed at facilities of the
Supercomputer Center of the Institute for Solid State Physics, the
University of Tokyo.


\vspace{1cm}




\begin{thebibliography}{}

\bibitem{dag94}E. Dagotto, Rev. Mod. Phys. 66, 763 (1994).
\bibitem{sca90}D. J. Scalapino, in {\em High Temperature Superconductivity-
the Los Alamos Symposium - 1989 Proceedings}, edited by K. S. Bedell,
D. Coffey, D. E. Deltzer, D. Pines, J. R. Schrieffer, (Addison-Wesley
Publ. Comp., Redwood City, 1990) p.314.
\bibitem{and97}P. W. Anderson, {\em The Theory of Superconductivity in
the High-T$_c$ Cuprates} (Princeton University Press, Princeton, 1997).
\bibitem{mor00}T. Moriya and K. Ueda, Adv. Phys. 49, 555 (2000).
\bibitem{hub63}J. Hubbard, Proc. Roy. Soc. London, Ser A 276, 238 (1963).
\bibitem{hir83}J. E. Hirsch, Phys. Rev. Lett. 51, 1900 (1983).
\bibitem{hir85}J. E. Hirsch, Phys. Rev. B31, 4403 (1985).
\bibitem{sor88}S. Sorella, E. Tosatti, S. Baroni, R. Car and M. Parrinell,
Int. J. Mod. Phys. B2, 993 (1988).
\bibitem{whi89}S. R. White, D. J. Scalapino, R. L. Sugar, E. Y. Loh, 
J. E. Gubernatis, and R. T. Scalettar, Phys. Rev. B40, 506 (1989).
\bibitem{ima89}M. Imada and Y. Hatsugai, J. Phys. Soc. Jpn. 58, 3752 (1989).
\bibitem{sor89}S. Sorella, S. Baroni, R. Car and M. Parrinello, Europhys.
Lett. 8, 663 (1989).
\bibitem{loh90}E. Y. Loh, J. E. Gubernatis, R. T. Scalettar, S. R. White,
D. J. Scalapino, and R. L. Sugar, Phys. Rev. B41, 9301 (1990).
\bibitem{mor91}A. Moreo, D. J. Scalapino, and E. Dagotto, Phys. Rev. B56,
11442 (1991).
\bibitem{fur92}N. Furukawa and M. Imada, J. Phys. Soc. Jpn. 61, 3331 (1992).
\bibitem{mor92}A. Moreo, Phys. Rev. B45, 5059 (1992).
\bibitem{fah91}S. Fahy and D. R. Hamann, Phys. Rev. B43, 765 (1991).
\bibitem{zha97}S. Zhang, J. Carlson and J. E. Gubernatis, Phys. Rev. B55, 
7464 (1997).
\bibitem{zha97b}S. Zhang, J. Carlson and J. E. Gubernatis, Phys. Rev. Lett. 78, 
4486 (1997).
\bibitem{kas01}T. Kashima and M. Imada, J. Phys. Soc. Jpn. 70, 2287 (2001).
\bibitem{yan98}T. Yanagisawa, S. Koike and K. Yamaji, J. Phys. Soc. Jpn. 
67, 3867 (1998).
\bibitem{yan07}T. Yanagisawa, Phys. Rev. B75, 224503 (2007).
(arXiv: 0707.1929)
\bibitem{yok87}H. Yokoyama and H. Shiba, J. Phys. Soc. Jpn. 56, 1490 (1987);
ibid. 56, 3582 (1987).
\bibitem{gro87}C. Gros, R. Joynt, and T. M. Rice, Phys. Rev. B36, 381 (1987).
\bibitem{nak97}T. Nakanishi, K. Yamaji and T. Yanagisawa, J. Phys. Soc. Jpn. 66, 
294 (1997).
\bibitem{yam98}K. Yamaji, T. Yanagisawa, T. Nakanishi and S. Koike,
 Physica C 304, 225 (1998); Physica B284, 415 (2000).
\bibitem{koi99}S. Koike, K. Yamaji, and T. Yanagisawa, J. Phys. Soc. Jpn. 68, 
1657 (1999); ibid 69, 2199 (2000).
\bibitem{yan01}T. Yanagisawa, S. Koike and K. Yamaji, Phys. Rev. B 64, 184509 (2001).
\bibitem{yan02}T. Yanagisawa, S. Koike and K. Yamaji, J. Phys.: Condens. Matter 
14, 21 (2002).
\bibitem{yan03}T. Yanagisawa, M. Miyazaki, S. Koikegami,  S. Koike, and K. Yamaji, Phys. Rev. 
B67, 132408 (2003). 
\bibitem{yan05}T. Yanagisawa, M. Miyazaki and K. Yamaji, J. Phys. Soc. Jpn. 78, 
013706 (2009).
\bibitem{miy04}M. Miyazaki, K. Yamaji and T. Yanagisawa, 
J. Phys. Soc. Jpn. 73, 1643 (2004).
\bibitem{luc12}L. T. Tocchio, F. Becca and C. Gros, Phys. Rev. B83, 195138 (2011).
\bibitem{yok12}H. Yokoyama, M. Ogata, Y. Tanaka, K. Kobayashi and
H. Tsuchiura, J. Phys. Soc. Jpn. 82, 014707 (2013).
\bibitem{miz86}T. Mizusaki, M. Honma and T. Otsuka, Phys. Rev. C53, 2786 (1986).
\bibitem{fei96}L. F. Feiner, J. H. Jefferson, R. Raimondi, Phys. Rev.
B53, 8751 (1996).
\bibitem{mai05}T. A. Maier, M. Jarrell, T. C. Schulthess, P. R. C. Kent,
and J. B. White, Phys. Rev. Lett. 95, 237001 (2005).
\bibitem{yan10}T. Yanagisawa, J. Phys. Soc. Jpn. 79, 063708 (2010).
\bibitem{sca86}D. J. Scalapino, E. Loh, and J. E. Hirsch, Phys. Rev.
B34, 8190 (1986)
\bibitem{bic89}N. E. Bickers, D. J. Scalapino, and S. R. White,
Phys. Rev. Lett. 62, 961 (1989).
\bibitem{hlu99}R. Hlubina, Phys. Rev. B59, 9600 (1999).
\bibitem{kon01}J. Kondo, J. Phys. Soc. Jpn. 70, 808 (2001).
\bibitem{yan08}T. Yanagisawa, New J. Phys. 10, 023014 (2008).
\bibitem{bla81}R. Blankenbecler, D. J. Scalapino, and R. L. Sugar, Phys. Rev.
D24, 2278 (1981).
\bibitem{sor01}S. Sorella, Phys. Rev. B 64, 024512 (2001).
\bibitem{noa97}R. M. Noack, N. Bulut, D. J. Scalapino and M. G. Zacher,
Phys. Rev. B56, 7162 (1997).
 
\end{thebibliography}
\end{document}